\documentclass[final]{elsarticle}

\usepackage{hyperref}
\usepackage{graphicx}%
\usepackage{multirow}%
\usepackage{amsmath,amssymb,amsfonts}%
\usepackage{amsthm}%
\usepackage{mathrsfs}%
\usepackage[title]{appendix}%
\usepackage{xcolor}%
\usepackage{textcomp}%
\usepackage{manyfoot}%
\usepackage{booktabs}%
\usepackage{algorithm}%
\usepackage{todonotes}%
\usepackage{algorithmicx}%
\usepackage{algpseudocode}%
\usepackage{listings}%
\usepackage[numbers]{natbib}

\journal{e-Informatica Software Engineering Journal}

\begin{document}

\begin{frontmatter}

\title{Challenges of Requirements Communication and Digital Assets Verification in Infrastructure Projects}


\author{Waleed Abdeen, Krzysztof Wnuk, Michael Unterkalmsteiner}
\address{Software Engineering, BTH, Karlskrona, Sweden}
\ead{first.last@bth.se}

\author{Alexandros Chirtoglou}
\address{HOCHTIEF ViCon GmbH, Essen, Germany}
\ead{alexandros.chirtoglou@hochtief.de}



\begin{abstract}
\emph{Background:} Poor communication of requirements between clients and suppliers contributes to project overruns,in both software and infrastructure projects. Existing literature offers limited insights into the communication challenges at this interface. \emph{Aim:} Our research aim to explore the processes and associated challenges with requirements activities that include client-supplier interaction and communication. \emph{Method:} we study requirements validation, communication, and digital asset verification processes through two case studies in the road and railway sectors, involving interviews with ten experts across three companies. \emph{Results:} We identify 13 challenges, along with their causes and consequences, and suggest solution areas from existing literature. \emph{Conclusion:} Interestingly, the challenges in infrastructure projects mirror those found in software engineering, highlighting a need for further research to validate potential solutions.
\end{abstract}


\end{frontmatter}


\section{Introduction}
\label{sec:introduction}
Developing software products and releasing them to the market is a complex process that requires efficient communication~\cite{hofmann_requirements_2001}. Inefficient communication manifests itself as communication gaps that lead to quality issues, wasted efforts, delays, and ultimately to failure to meet the customers' expectations \cite{bjarnason2011requirements}. Moreover, inefficient use of requirements artifacts may impede requirements communication in software projects~\cite{liskin2015artifacts}. What is common for large software and infrastructure projects is their complexity that often implies the use of suppliers \cite{Brereton} to deliver significant parts of the solution, e.g. automotive industry often uses suppliers to develop software for cars. 

The friction-free economic principles of software industry help involve suppliers from geographically distant locations, primarily for the cost reduction and missing competence availability. One of the main challenges in involving suppliers is establishing good communication principles, especially agreed quality levels, and also communicating potential difficulties and issues early ~\cite{EbertSuppliers}. Moreover, Vullinghs et al.~\cite{vullinghs2000experiences} highlighted the need to derive supplier process requirements for software quality assurance processes between Daimler-Chrysler and its suppliers. Zheng et al. ~\cite{zheng2020inter} discovered that the uncertainty of the requirements increases the negative influence of coordination efforts on trust for automotive new product development projects. Jiang et al.~\cite{JIANG2018925} discovered that both goal interdependence and resource
interdependence have a positive influence on coordination and cooperation in new software development programs that use external partners (suppliers).

The ability to communicate clients' needs and requirements to suppliers is a key success criterion in infrastructure projects~\cite{songer1997project,ibbs2003project, kania_communication_2020}. Moreover, effective communication facilitates stakeholder engagement during the change management process in infrastructure projects~\cite{BUTT20161579} and helps to clarify realistic stakeholder expectations, while insufficient communication drives uncontrolled change and can contribute to project delays~\cite{zhao2010prediction}.



Infrastructure projects often last decades in planning and execution~\cite{BUTT20161579}, and involve a large number of specialized project teams (e.g., owners, contractors, consultants, designers) with specific core competencies. These projects are also featured with uncertainty,  fragmentation~\cite{WU20171466}, and high complexity and inter-organizational task interdependence, which makes communication ever more important~\cite{badir2012conceptual}. Therefore, it is crucial to understand which challenges arise in client-supplier communication in design-build projects, in particular since contracts of construction projects are inherently incomplete~\cite{demirel2017flexibility}.  

Several studies have explored the challenges associated with general communication in large construction projects~\cite{malik_exploring_2021}, requirements allocation~\cite{vermillion_investigation_2020}, verification and validation~\cite{makkinga_successful_2018},  system integration~\cite{madni_systems_2014}, the relationship between the communication-conflict interaction and project success~\cite{WU20171466}, trust between firms and suppliers~\cite{XU202132}, encoding and decoding communication competencies in project management~\cite{HENDERSON2004469}, or inter-cultural communication~\cite{LOOSEMORE199995}. 

However, little research effort was dedicated to exploring the processes and associated challenges with requirements activities that include client-supplier interaction and communication in infrastructure projects. Some work exists that focuses on requirements, for example Zheng et al.~\cite{zheng2020inter} discovered that requirements uncertainty increases the negative influence of coordination efforts on trust with suppliers . Fricker et al. \cite{fricker2009handshaking} suggested handshaking with implementation proposals that could be used to improve the communication between the client and supplier, however no empirical evidence is provided in this case. In this work. we are interested specifically in the requirements verification process that ensures the conformity of the product or any parts of it to what is required, and in the requirements validation process that ensures that the product delivers what the client actually needs~\cite{Incose2015Incose}.

This study is part of a research project that aims to investigate methods that can improve functional requirements traceability at different stages in construction projects. Functional requirements should be linked to a digital twin~\cite{semeraro2021digital}, a collection of digital assets that represent the facility and are linked to the physical facility. Digital assets are defined as any digital file (textual, images, audio, or video) stored on any electronic device (e.g., computer, mobile phone, or cloud) with the right to own the file~\cite{toygar_new_2013}.
In infrastructure projects, Computer-Aided Design models (CAD), Building Information Models (BIM), and System Information Models (SIM)~\cite{love_systems_2016} are typical digital assets.
Traceability between functional requirements and digital assets would allow for efficient and effective requirements verification before construction has started, enable monitoring during construction, as well as make it possible to follow-up on the original requirements during the decades of operation and maintenance.

We frame our research using the design science process, which aims to extend knowledge by identifying and addressing specific problems in practice. This study explores the nature of client-supplier requirements communication in large infrastructure design-build projects, where the supplier is responsible for the design, construction, and requirements validation and verification processes. Such a description can help to understand the challenges that practitioners encounter and either point to solutions that exist in the literature or describe a research gap.

We conducted two case studies where we explored the client-supplier interface. In the two projects, Anonymous acted as client and two other companies acted as contractor and sub-contractor. The semi-structured interviews were designed to understand \emph{how} requirements communication, requirements validation, and the verification of deliverables are conducted. We transcribed the audio recordings and coded the data to answer the research questions.

We present an overview of the client-supplier interface and develop process diagrams for each investigated case. Furthermore, we identify and describe 13 challenges faced by the managers and engineers working in these processes. Finally, we map these challenges to potential solutions from the system and software engineering domain. \emph{Identify conflicts early}, \emph{requirements abstraction}, \emph{requirements are impossible to build} and \emph{granularity of traces}, are challenges that are also faced in software engineering domain and are important to address.

The remainder of this paper is structured as follows. We present in Section~\ref{sec:related work} literature that is related to our work. In Section~\ref{sec:methodology}, we present our research methodology and study design. We present our research results in Section~\ref{sec:results} and discuss them in Section~\ref{sec:discussion}. In Section~\ref{sec:conclusion}, we conclude and outline future work.

\section{Related Work}
\label{sec:related work}

The focus of our investigation is the processes related to requirements engineering on the interface between clients and suppliers in the context of construction projects. In this section, we discuss, therefore, related work from systems engineering and communication in large construction project research areas.

\subsection{Communication in Software Engineering}

Bjarnason et al. investigated geographical, cognitive and psychological distances in communication between teams \cite{bjarnason2022inter}. Liskin investigated how  artifacts support or impede requirements communication \cite{liskin2015artifacts}. Liebel et al. studied communication problems in automotive requirements engineering, concluding that it is important to establish communication channels outside the fixed organisation structure \cite{liebel2018organisation} . Iqbal et al. developed a framework to address communication issues during
requirements engineering process for software development
outsourcing \cite{iqbal2018framework}. Fricker et al. \cite{fricker2009handshaking} suggested handshaking with implementation proposals that could be used to improve the communication between the client and the supplier. Pernstal et al.~\cite{pernstaal2015requirements}
developed the BRASS framework to support coordination and communication of inter-departmental requirements in the large-scale development of software-intensive systems . However, the framework does not focus on requirements validation, but rather on supporting the selection of the most suitable implementation plan.  

\subsection{System Engineering}

RE and verification are of high interest to the system engineering community, which can be seen by the several conducted studies covering different system engineering processes and their challenges~\cite{madni_systems_2014,maalem_challenge_2016,vermillion_investigation_2020,lynghaug_investigating_2022,de_graaf_level_2023}.

Madni et al.~\cite{madni_systems_2014} explored the challenges in system integration in the defense and aerospace domain. They propose an ontology-based system integration approach to facilitate communication between the different stakeholders.
Makkinga et al.~\cite{makkinga_successful_2018} conducted semi-structured interviews in three mid-sized projects in the Netherlands. Their focus was on the verification and validation problems in construction projects and possible solutions to these problems. They advise more research work in the validation process. Vermillion et al.~\cite{vermillion_investigation_2020} compare the requirements allocation and objective allocation in projects with outsourced design. They propose an approach for requirements allocation for the design process. Moreover, they see an asymmetry in knowledge between the clients who outsource the design and agents designing the system, which needs to be addressed to reduce negotiation iterations.

System engineering processes and practices in the construction domain have also been of interest to researchers~\cite{lynghaug_investigating_2022,de_graaf_level_2023}.
Lynghaug et al.~\cite{lynghaug_investigating_2022} explored the state of system engineering practices in the Norwegian construction domain. They conclude by providing recommendations on implementing and improving system engineering practices in the construction domain. One of the most critical system engineering processes is requirements analysis. Raatikainen et al.~\cite{raatikainen2011challenges} highlighted the challenge of efficient  communication and management of requirements in the nuclear energy domain. De Graaf et al.~\cite{de_graaf_level_2023} investigated the state of practice of six cases of sub-contracted design work in civil engineering projects. They identified three factors that affect the sub-contracted work, mainly: Building Information Model (BIM) interoperability, time pressure, and employee availability.

\subsection{Communication in Large Construction Projects}

Malik and Taqi explored the relationship between communication and the success of construction projects~\cite{malik_exploring_2021}. They found that process conflict and relationship conflict have a negative impact on communication and project success. Saxena and McDonagh  have investigated communication breakdowns and concluded that change management requires a multilevel communication approach~\cite{SAXENA2022181}. Daim et al. have also looked into communication breakdowns among global virtual teams, and these breakdowns tend to threaten project delivery. The authors found five factors impacting communication breakdowns: trust, interpersonal relations, cultural differences, leadership and technology ~\cite{DAIM2012199}.

Wu et al. investigated the relationship between the communication-conflict interaction and project success~\cite{WU20171466}. Butt et al. have looked into how effective communication facilitates stakeholder engagement during the change management process and changing project culture~\cite{BUTT20161579}. Communication is a powerful tool in ensuring participation in change management processes. Lack of communication leads to teams focusing on task performance and efficiency rather than empowerment and involvement. Henderson et al. looked into the impact of communication norms on global project teams and the individuals' project satisfaction and performance~\cite{HENDERSON20161717}. Henderson also looked into encoding and decoding communication competencies in project management~\cite{HENDERSON2004469}. Loosemore and  Muslmani investigated inter-cultural communication challenges in Persian Gulf projects, highlighting the need for a better understanding of cultural diversity~\cite{LOOSEMORE199995}. Similar recommendations were reported by Ochieng and Price, who studied the cultural variation of project managers in Kenya and the UK in communicating effectively on multicultural projects~\cite{OCHIENG2010449}.

\subsection{Research Gap}
Requirements validation, communication and verification are important in avoiding project failures~\cite{terry2005requirements,kania_communication_2020}. While researchers from the software engineering domain have investigated the area~\cite{bjarnason2014challenges,hotomski_exploratory_2016}, it remains greatly unexplored or neglected in the system engineering domain.
Hence, due to the lack of studies that investigate requirements validation, requirements communication and digital assets verification in the system engineering domain, we are conducting this study to fill this gap.


\section{Methodology}
\label{sec:methodology}
Due to the aim of the research project as a whole, which is to identify and address problems related to the requirements communication process and the traceability of functional requirements to the digital twin, we frame our research as design science problem~\cite{hevner2004design, wieringa2014design}. Hevner~\cite{hevner2007three} has depicted design science as a process intertwining relevance, rigor, and design cycles. In the study reported in this paper, we investigate the challenges of client-supplier communication in system engineering projects; in other words, we identify the \emph{relevance} of the problem in the infrastructure domain. Thereafter, we plan to design a solution for one of the problems we identify in this study, apply it to the problem, and further improve it (\emph{design}). Finally, we plan to verify that the solution fits the problem through implementation to other problem instances from the software engineering domain (\emph{rigor}). In this study, we focus only on determining the \emph{relevance} of the problem.

\begin{table*}[htb]
\centering
\caption{Research Questions}
\label{tab:research questions}
\footnotesize
\begin{tabular}{lp{0.35\textwidth}p{0.50\textwidth}l}
\toprule
\textbf{Id} & \textbf{Research Question} & \textbf{Motivation} \\
\midrule
    RQ1
    &
    What are the practices and challenges in system requirements validation? 
    & 
    One of the main documents that are communicated between the client and the supplier in design-build contracts are system requirements. Hence it is important to have those requirements validated. We explore the system requirements validation process and the challenges associated with it.
    \\
    RQ2
    &
    What are the practices and challenges in system requirements communication between the client and the supplier?  
    &
    Explore the requirements communication process between the client and the supplier, and what different formats the requirements take before they are translated into a design. Moreover, we want to explore the difficulties in the process and opportunities for improvements.
    \\
    RQ3
    &
    What are the practices and challenges in digital assets verification?  
    &
    Explore any quality checks done by the client or the supplier on the design documents and any challenges associated with the process.
    \\
\hline  
\end{tabular}
\end{table*}

We designed the study according to the case study guidelines by Runeson et al.~\cite{runeson2009guidelines}, who advocate for defining a detailed case study protocol that reflects the changes made during the iterative process of data collection and analysis.
The research questions with the motivation and alignment to our objective are listed in Table~\ref{tab:research questions}.

\subsection{Case description}
The two cases were selected from \emph{Anonymous'} projects based on availability and access to information. Both projects distinguish between project-specific requirements and regulatory requirements. Project-specific requirements describe needs that pertain to the particular facility, originating from diverse stakeholders such as the government, communes, and land owners. Regulatory requirements define needs that are relevant to all applicable facilities. While the client specifies the project-specific requirements, it is the responsibility of the supplier to identify and comply with the relevant regulatory requirements for the designed facility.

Case One is a road project that includes the design and building of roads and bridges for cars, pedestrians, and cyclists. At the time of the study (April-June 2020), the project was in construction (3 years duration). Case One, with a budget of 35M USD, is part of a larger road project with an estimated duration of 30 years (15 planning, 15 construction) and a budget of 4B USD. Case Two has specified approximately 700 project-specific requirements and is associated with 12.000 regulatory requirements.

Case Two is a railway project with the design and build of a high-speed train connection, including rail, bridges, and tunnels. The project was in its early stages (requirements specification and initial design) at the time of the study. The estimated project duration for Case Two is 34 years (9 pre-study, 10 planning, 15 construction) and a budget of 8.8B USD. Case Two has specified approximately 1.000 project-specific requirements and is associated with approximately 15.000 regulatory requirements\footnote{In both projects, the number of project-specific requirements varied over time. Furthermore, the number of associated regulatory requirements is an estimate by a requirements engineering lead at \emph{Anonymous}. Determining the exact numbers would require that every contractor keeps track of the regulations they need to comply to, which is, as we shall see, not the case.}.

\begin{table*}[bt]
\centering
\caption{Companies Overview}
\label{tab:companies overview}
\resizebox{1\textwidth}{!}{
    \begin{tabular}{llllll}
    \toprule
    \textbf{Company} & \textbf{Industry} & \textbf{Role} & \textbf{Case} &\textbf{Company Size} \\
    \midrule
    
    A & Construction & Contractor & Case One & 45000 \\
    B & Design & Sub-Contractor & Case One & 6000 \\
    C & Transportation and Infrastructure & Client & Case One, Case Two  & 9400  \\
    
    \hline 
    \end{tabular}
}
\vspace{1ex}

{\raggedright The unit of measure for company size is the average number of employees \par}

\end{table*}

\subsection{Data collection}
\label{sec:data collection}

\begin{table}[bt]
\caption{Interviewees Roles}
\label{tab:interviewees roles}
\footnotesize
\begin{tabular}{lp{0.40\textwidth}llll}
\toprule
\textbf{Id} & \textbf{Role} & \textbf{Company} & \textbf{Experience} \\
\midrule
A1 & Design Manager & A & 20+ years \\
A2 & Tender Manager & A & 10 years \\
A3 & BIM Manager & A & 7 years \\
A4 & Design Manager & A & 10 years \\
B1 & Discipline Leader & B & 15 years \\
B2 & Head of Design & B & 25 years \\
C1 & BIM Specialist & C & 9 years \\
C2 & Requirements Specialist & C & 11 years \\
C3 & Requirements Engineer & C & 8 years \\
C4 & Head of Tech and Environment & C & 24 years \\

\hline  
\end{tabular}

\vspace{1ex}

{\raggedright The experience presented in this table is the overall industry experience. \par}

\end{table}

We conducted semi-structured interviews with the participants from all three companies listed in Table~\ref{tab:companies overview}. Semi-structured interviews are fit for a study where a clear hypothesis does not exist and the research questions are of explorative nature~\cite{runeson2009guidelines}. In our case, we explored the system engineering processes that involve client-supplier communication, but we don't know to what extent these processes were implemented. 

We formulated the interview questions in three main themes, according to our stated research questions: \textit{requirements validation}, \textit{requirements communication}, and \textit{digital assets verification}. Then we adapted the interview questions to each interviewee role. For example, when we interviewed a requirements engineer, we focused our questions on \textit{requirements validation} and \textit{requirements verification}.

We interviewed ten people from three companies. Table~\ref{tab:interviewees roles} shows their roles, companies, and industry experience. The interviewees had between 7 and 24 years of experience and worked in roles related to requirements engineering, design, and project management. Convenience sampling was used when choosing the participants for the interviews. We asked our company contacts for people who work with requirements and system design, then they suggested people based on their availability. In this paper, we use the acronym XY to reference interviewees, where X is a letter referencing the company [A,B,C] and Y is a number referencing an interviewee in that company.

Up to three researchers attended the interviews (observer triangulation). The first author led the interview by asking questions, and the other two authors observed and asked follow-up questions if necessary. We conducted each interview as follows:
\begin{enumerate}
\item A week before the interview, we sent the informed consent letter to the participants and asked them to return a signed copy within two weeks after the interview.
\item Before the interview started, we briefly described the purpose of our study, and the expected outcome and explained how we are going to conduct the interview.
\item We started recording the interview and asked the questions.
\end{enumerate}

After we finished conducting the interviews, the first author transcribed them. The third author listened to the interviews' recordings to verify the transcript. We sent each transcript out to the interviewees for comments. We received corrections from one participant, clarifying statements that were incomplete due to recording issues caused by the recording equipment.

\subsection{Data analysis}
\label{sec:data analysis}

\begin{figure}
    \centering
    \includegraphics[width=0.7\textwidth]{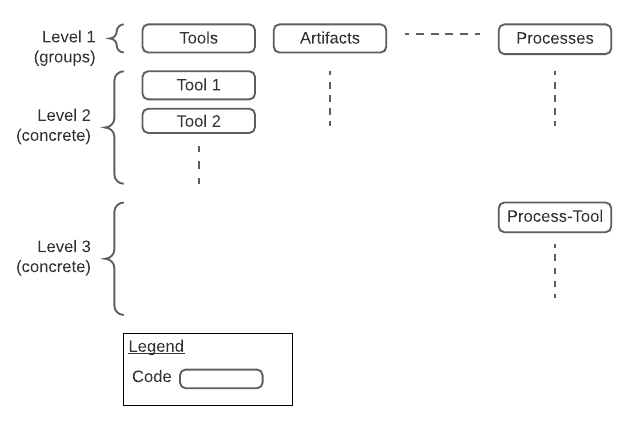}
    \caption{Coding Levels}
    \label{fig:coding}
\end{figure}

We coded the interview transcripts with an initial set of codes that the first and third author created in alignment with the research questions. The first author quoted and labeled parts of the transcript using these codes. Feedback sessions were conducted between the first and third author to refine the codes. Several coding iterations were done until we had enough data to answer the research questions. A final verification of the quoted text was done by the third author. The codes were created on different levels of abstraction, as seen in Figure~\ref{fig:coding}. This helped us to use relevant codes when answering the research questions, as exemplified next.

\paragraph{Coding example} When coding the answers to the interview questions related to the requirements validation process, we started with the codes: 1) \textit{requirements validation process} to quote the activities associated with the requirements validation process and 2) \textit{requirements validation challenge}, to quote the challenges of that process. These codes were part of the group \textit{requirements validation}. Then we added more codes based on the content, e.g., \textit{requirements documentation} and \textit{cost estimation} to quote the artifacts used in the validation process. After that, we created codes on a higher abstraction level e.g., \textit{tools}, \textit{artifacts}, and \textit{processes}.

Using this way of coding resulted in codes on different abstraction levels. This helped retrieve all information in a whole area, e.g., the requirements validation process, and in detail in that process, e.g. actors in the requirements validation process.

\subsection{Threats to validity}

We use the framework by Runesson et al.~\cite{runeson2009guidelines} to discuss the validity threats of our study. The framework lists four categories of case study validity: construct, internal, external, and reliability. We analyzed the validity threats of our study from the beginning of our study, and we continued to revise these threats to minimize them.

\paragraph{Construct Validity} We mitigated threats to design and execution by involving more than one researcher in designing and conducting the study. The involved researchers reviewed the research questions and the case study protocol. Feedback sessions were conducted to discuss the protocol and make improvements. For example, the first author wrote a set of interview questions per studied process. During a feedback session, the third author suggested that the questions should be adapted to each interviewee's role. On that premise, the interview questions were adapted to each role by the first author, and the other researchers reviewed them. The improvement of the interview questions for each interview was done by the researchers before the start of said interview.

\paragraph{Internal Validity} Although the involved researchers revised and improved the case study protocol, there might be a bias or errors in the collected data. To reduce the risk of bias in data collection, between two and three researchers were present during each interview. One researcher led the interview, and the other researchers observed the interview and asked follow-up questions if necessary. In addition, after an interview was completed, we transcribed the interview, a second researcher verified the transcript and finally sent it out to the interviewee for comments.

Another internal threat to validity is the risk of bias when coding the transcripts as part of the data analysis phase. We mitigate this threat by assigning the coding task to one researcher and the verification of those codes to another. Eventual disagreements were resolved in meetings.

\paragraph{External Validity} It could be argued that generalization would be difficult with two case studies and that more case studies may be required to increase the validity of the study and achieve generalization. However, generalization can be achieved by individual cases. It is a matter of the selected cases. As argued by Flyvbjerg, one of the five misunderstandings of a case study is the inability to generalize from a single case~\cite{flyvbjerg_five_2006}. We focus on analytical generalization rather than statistical generalization by providing detailed case descriptions and discussing the implications of our findings.  Moreover, we believe that our cases are a good representation of the population (infrastructure projects). They are two large infrastructure projects, with large-sized (45000+ employees) companies involved. Case One is in its final stages, while Case Two is in its early stages.

\paragraph{Reliability} To ensure that our study is repeatable, we present in Section~\ref{sec:methodology} the protocol of our study. We explain the setup of our studied cases, list the cluster of the interview questions, present the steps in which we conducted the interviews, and the coding mechanism of the interview data.

\section{Results}
\label{sec:results}

We present the results in three sections following the themes of the interview questions: requirements validation, requirements communication, and digital assets verification. In each section, we describe the process and the challenges faced in that process. A challenge is presented as the description, causes, and consequences of that challenge. Some challenges have no causes or/and consequences since we could not identify them from the interviews. We illustrate the results in figures that show the process flow, artifacts, and actors. Each figure is divided into columns representing the parties involved in the project: client, contractor, or subcontractor. In some figures, there is a question mark which is a placeholder for information that was not clear from the interviews. 

\subsection{Requirements validation}

\begin{figure*}
    \centering
    \includegraphics[width=\textwidth]{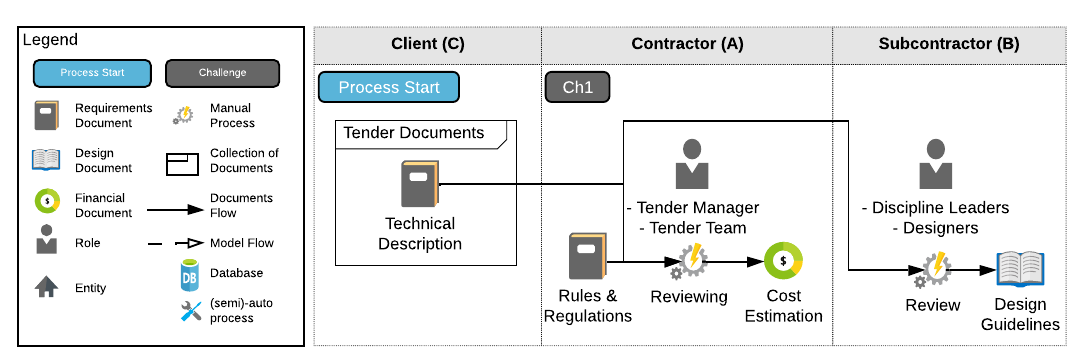}
    \caption{Requirements Validation - Case One (legend applies to all figures)}
    \label{fig:rv-case1}
\end{figure*}

Figure~\ref{fig:rv-case1} depicts the requirements validation process in Case One. At the contractor, the tender manager and their team conduct a review process. The inputs to the process are the technical description and the general regulations (local to the country) that apply to the project. The output of the process is a cost estimate spreadsheet, which estimates the project cost and is used when preparing the contract for the project. Another requirements review process is conducted at the subcontractor by the discipline leaders and designers. They take as input the technical description from the contractor and the general regulations. The output of this process is the design guidelines document that helps designing the models for the project. 

\begin{figure*}
    \centering
    \includegraphics[width=\textwidth]{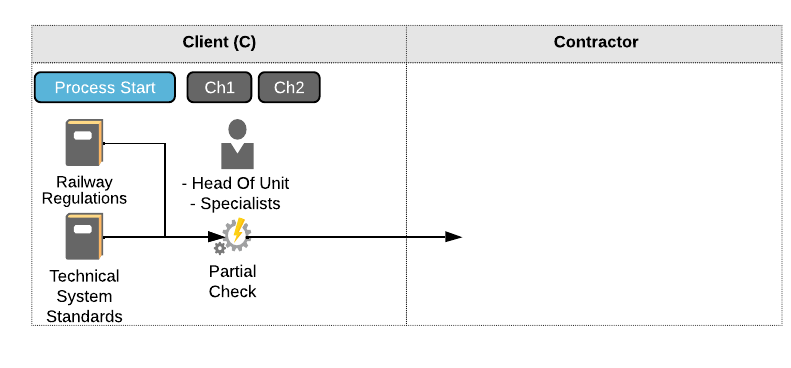}
    \caption{Requirements Validation - Case Two}
    \label{fig:rv-case2}
\end{figure*}

Figure~\ref{fig:rv-case2} shows the requirements validation process in Case Two. The client conducts a partial check on the requirements. The inputs to this check are the railway regulations and technical system standards and the outputs are the validated requirements. We do not have information about the requirements validation process on the contractor side, because the people we interviewed in Case Two are from the client-side.

We have identified the following two challenges that are faced by the roles involved in the requirements validation process.

\subsubsection{Challenge 1: Prioritizing the requirements validation process}
\label{sec:RVC1}
Prioritizing the requirements validation process is a challenge faced by the contractor in Case One and the client in Case Two. 
The time to validate the requirements by the contractor is seen to be short in Case One, as reported by the tender manager (A2). When the tender project is announced, the contractor has to respond with the cost estimate in a matter of weeks. This makes the validation of the requirements challenging for the contractor who has to do it in a short time. ``The client spends four years coming up with the design (presented by the requirements), and we have to price it in four or eight weeks, coupled with a tender model that is purely price-driven. It is a very unhealthy situation'' (A2).
The requirements validation process requires experience and knowledge in the discipline, which makes the prioritization of this process and finding the right people to perform it difficult for the client. ``When you talk about validating requirements, it is not our top priority'' (C2).

\textit{Ambiguous and conflicting requirements} is a cause of this challenge in Case One. The requirements validation process could take more than a couple of months when the supplier finds requirements that need clarification or revision by the client. In Case Two, \textit{lack of resources} is seen as a cause of this challenge. As C2 said, ``there are not enough railway specialists in the company.'' The people involved in Case Two are involved in other projects and have other responsibilities at the client, making it difficult to find the right people to validate the requirements. 

\textit{The requirements validation process is overlooked} is seen by the client as a consequence of not prioritizing the said process. Since it is not always feasible to allocate resources for the validation process, it gets less priority over other activities, leading to the process being skipped in many cases.
Furthermore, the contractor sees \textit{added responsibility} as a consequence. Since the contractor submits the bid to the project based on the requirements, they take the responsibility if they have not validated the requirements properly: ``Do not put all the risk on us if we can't find the problem in such a short time'' (A2).

\subsubsection{Challenge 2: Identify conflicts early}
It could be difficult to identify conflicts in the requirements early as seen in Case Two by the requirements engineer (C3) and head of environment (C4). Although a partial check on the requirements is conducted at the client side, some conflicts may go undetected until later at the design stage. For example, when verifying tunnel-related requirements, one (requirements engineers or designers) may not be able to identify water pipes intersection with an obstacle due to the land geometry. ``It is hard to see that before when you haven't drawn the lines for the pipes or the tunnel'' (C3).

The main cause of this challenge is not clear. When we asked C3 whether the lack of information or having too much information is the cause of difficulties identifying those conflicts, they answered ``it depends'' (C3).


\subsection{Requirements communication}

\begin{figure*}
    \centering
    \includegraphics[width=\textwidth]{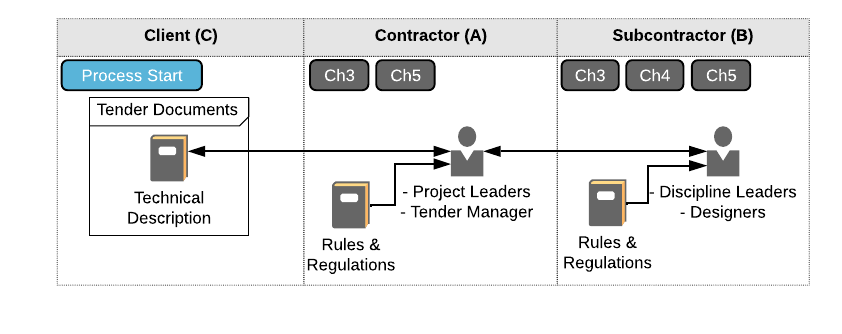}
    \caption{Requirements Communication - Case One}
    \label{fig:rc-case1}
\end{figure*}

Figure~\ref{fig:rc-case1} shows the requirements communication process in Case One. The project's technical requirements, called technical description document, are a part of the tender documents. The project leaders and tender manager on the contractor side get the technical description from the client, and then they communicate these documents to the discipline leaders and designers on the subcontractor side. Additional requirements from the rules and regulations apply to the project in Case One. The contractor and subcontractor are responsible for finding those additional requirements that apply to the project. 

\begin{figure*}
    \centering
    \includegraphics[width=\textwidth]{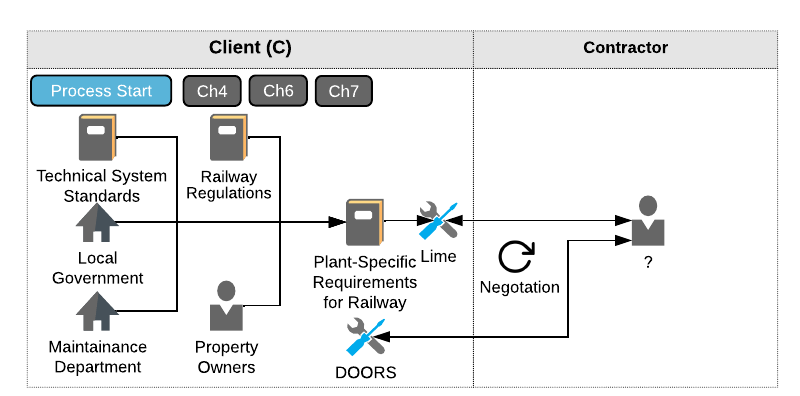}
    \caption{Requirements Communication - Case Two}
    \label{fig:rc-case2}
\end{figure*}

Figure~\ref{fig:rc-case2} shows the requirements communication process in Case Two. In this case, there are three main types of requirements sources: 1) documents, e.g., technical system standards, or railway regulations; 2) internal or external departments, e.g., local government or maintenance department; and 3) people who reside or work in the project area and who could be affected by the project outcome, e.g., property owners. The client compiles the requirements from all the mentioned sources into project-specific requirements for the railway. Those requirements are stored in a software called LIME, which is used to negotiate requirements between the client and the contractor. The documents stored in LIME are living documents, where both the client and supplier communicate the requirements and their feedback through an iterative process. After the client and the contractor have agreed on the requirements, those requirements are stored in DOORS\footnote{requirements management software}.

The interviewees reported five challenges in relation to requirements communication.

\subsubsection{Challenge 3: Misinterpretation of requirements}
\label{miss-interpretation of requirements}
All parties working on the project have access to the same requirements documents, however, each party has their interpretation that could differ from the other parties' interpretation. This was reported in both cases by the design manager (A1), tender manager (A2), and discipline leader (B1).
Moreover, the requirements can be unclear or difficult to understand. As one interviewee on the subcontractor side mentioned, ``we have had a whole lot of discussions regarding these requirements in the technical description, and it is constantly not crystal clear, so it is very much up to interpretation'' (B1). Also, the client thinks that they could improve their writing; as C4 said, ``we often think that we are quite clear in our communication, but often it's not the case'' (C4).

\textit{Lack of knowledge in local projects} and \textit{requirements are open for interpretation} are of the causes of this challenge, as seen by the interviewees. Some people working on the project may lack experience in projects from the this specific country, which could lead to a different interpretation of the requirements. ``If you have any country A engineer and you want to make a bridge design, his background is slightly or completely different from the country B engineer'' (A2). The subcontractor sees it as one of the causes of misinterpretation of requirements as put by one interviewee ``There is never really a correct answer. There is not just one solution that you can do'' (B1). 

The misinterpretation of requirements may lead to rework and extra costs. If the subcontractor's interpretation is different from the client's interpretation, then additional work should be done by the subcontractor. As B1 said, ``if our interpretation of the requirements is not that what the client wants us to do, then there is some additional work for us'' (B1). In special cases, the misinterpretation of requirements could lead to extra costs paid by the contractor, as confirmed by A1, ``I was not informed at all what a requirement (temporary use of land for the project) means''.

\subsubsection{Challenge 4: Long time to communicate requirements (changes and questions)}
The requirements communication process between the different parties (client-supplier) takes a long time. This process includes requirements change requests and any requirements-related questions raised by the contractor or subcontractor. This challenge was seen in Case One and Case Two by the discipline leader (B1) and the head of environment (C4). 
All requirements-related communications between the client and subcontractor go through the contractor, which takes a long time: ``it could take quite a bit of time'' (B1). The client also sees that the change process takes a long time from the contractor side as well. C4 gave an example of a case, ``one occasion we did a very big change of the requisition of the whole project ... but then it took like a year for the consultants to respond and tell us how the changes would be interpreted and implemented'' (C4).

The interviewees did not explicitly identify the causes or consequences of this challenge.

\subsubsection{Challenge 5: Finding the correct information}
The documents communicated by the client do not include all the requirements that apply to the project, and it is challenging to find all requirements that apply by the contractor and subcontractor. This challenge was reported in Case One by the design manager (A4) and head of design (B2). As seen in Figure~\ref{fig:rc-case1} and Figure~\ref{fig:rc-case2}, there are additional requirements documents that the client does not communicate as part of the tender documents. Those documents are local rules and regulations. The client only refers to those documents, and the contractor and subcontractor need to find those documents. ``The tricky task to make sure you have all requirements that have to be followed'' (A4).

A cause of this challenge is \textit{Lack of knowledge in local projects}, similar to that of miss-interpretation of requirements challenge explained in Section~\ref{miss-interpretation of requirements}. Some people with different background could lack experience in projects in a specific country. In this case, people with experience in local projects are consulted to find the right information.


\subsubsection{Challenge 6: Requirements elicitation and validation with non-technical stakeholders}
The requirements elicitation and validation process is a challenging task since many stakeholders are involved. This challenge was reported in Case Two by a requirements engineer (C3). The client elicits requirements from many sources and stakeholders, as seen in Figure~\ref{fig:rc-case2}. Therefore, the way the client communicates the requirements needs to be adapted to the audience. ``We can have a super model to communicate with the supplier, but we can't use that one when we meet the restaurant owner in the city'' (C3).

The cause of seeing the requirements elicitation and validation process as a challenging task is because there are \textit{different parties and many stakeholders with different backgrounds} involved in the process.

This challenge adds additional work to the client, as the client needs to adapt the requirements to the audience. For example, sketches and drawings need to be done so the client can communicate the requirements with the property owners.

\subsubsection{Challenge 7: Requirements abstraction}
The requirements communicated by the client would likely have variations in abstraction levels. The challenging part is to find the right level of granularity for these requirements. This is reported in Case Two by the head of environment (C4). A too specific requirement could constrain the supplier in developing the solution, ``sometimes we were getting the wrong answer when we are too specific in demands'' (C4), and a too abstract requirement is open for miss interpretation ``often we have to explain more to make the requirements more specific'' (C4).

We did not identify clear causes or consequences for this challenge from the interviews.

\subsection{Digital assets verification}

\begin{figure*}
    \centering
    \includegraphics[width=1\textwidth]{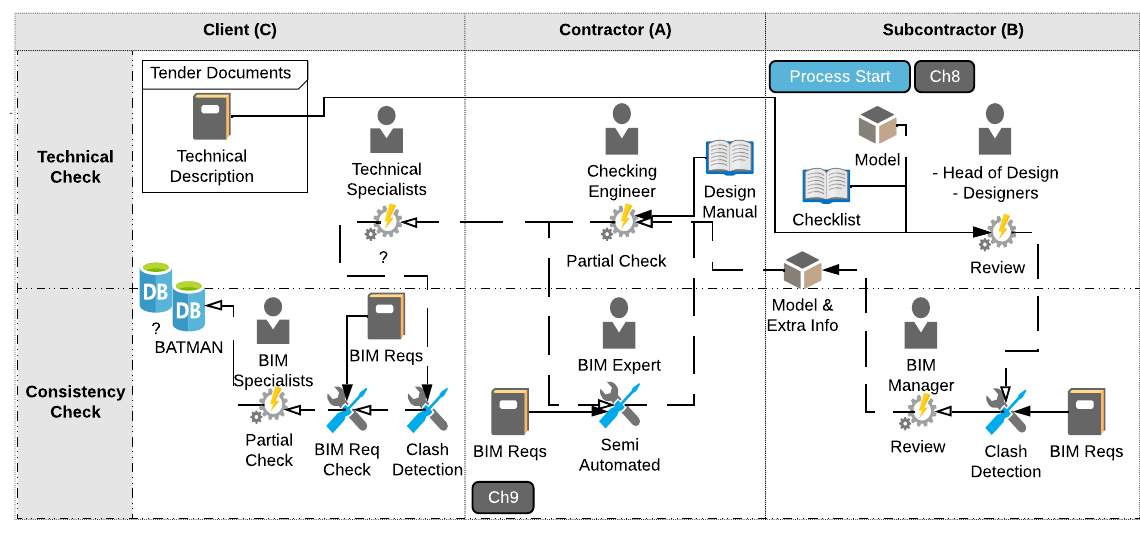}
    \caption{Digital Assets Verification - Case One}
    \label{fig:dav-case1}
\end{figure*}

Figure~\ref{fig:dav-case1} shows the digital assets verification process in Case One. There are two verification processes present in the figure. The technical check is the process of verifying the produced digital assets' conformity with the specified technical requirements. The consistency check is the process of verifying the conformity of digital assets with the BIM requirements. BIM requirements specifies how the design models should be delivered (e.g., the file's extension to be used for delivering a model, or the units of measure used).

As the subcontractor designers produce the model, a review process is conducted by those designers and the head of design. The process takes as input 1) the produced model, which could be 2D drawings/3D models/BIM, 2) a checklist that is prepared by the designers based on the technical requirements before the implementation, 3) the technical requirements of the project. The output model of this process, along with BIM requirements, goes into a clash detection tool, which checks whether the models have any conflict in design objects. After that, a manual review process is made by the BIM manager. 

At the contractor, a similar verification process on the model is followed. First, a technical check is done with the design manual, which is produced at the beginning of the project. This check is done by a checking engineer and is mainly based on experience with verification of similar models. The model is sent to the BIM expert who conducts pre-configurated semi-automated checks on the attributes of the BIM objects based on the client's BIM requirements, using internal tools with customizable configuration possibilities. The technical and consistency check is an iterative process. 

Then the model is sent to the client for approval. At the client, there are technical specialists who do their own technical checks while taking technical requirements as input. After that, a series of consistency checks are conducted: clash detection, BIM requirements check, and partial check by BIM specialists. When the model passes all those checks, it is stored in the appropriate database.

\begin{figure*}
    \centering
    \includegraphics[width=\textwidth]{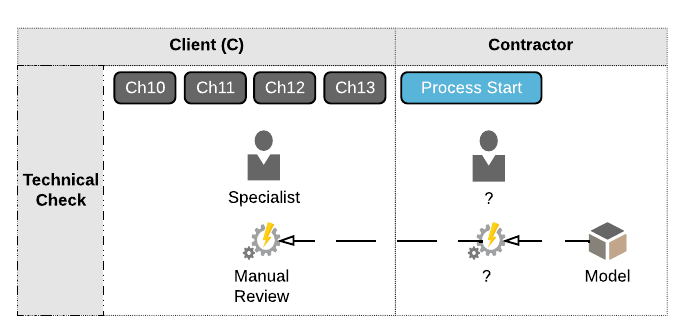}
    \caption{Digital Assets Verification - Case Two}
    \label{fig:dav-case2}
\end{figure*}

Figure~\ref{fig:dav-case2} presents the digital assets verification process in Case Two. When the models are delivered to the client for approval, the specialist in the domain conducts a manual review process to verify those models. We also know that there is some kind of check being done by the contractor.

We have identified six challenges related to the process of digital assets verification.

\subsubsection{Challenge 8: Requirements are impossible to build}
In some cases, during the verification of digital assets, engineers detect requirements that are impossible to build. This challenge was reported in Case One by the discipline leader (B1). The contractor and subcontractor are limited by what they can change in the requirements. Then during the verification of digital assets, some requirements appear to be impossible to build. ``The solution that company C stated simply can't be done'' (B1). Although the client may be validating the requirements, it is still the contractor's and subcontractor's responsibility to make sure those requirements are sound and possible to build.

\textit{The client specifies solutions rather than requirements} were the cause of this challenge. The requirements provided by the client do not have much room for the contractor or subcontractor to come up with a solution. Rather the requirements specify an actual solution that the contractor and subcontractor need to follow. Therefore, if those requirements/solutions have conflicts, they will likely show during the verification process (since they might be overlooked in the requirements validation process, as discussed in Section~\ref{sec:RVC1}.

The consequence of this challenge is \textit{extra effort spent on rework} to be done by the subcontractor. During the verification process, if some requirements were detected to be impossible to build, then the subcontractor needs to come up with a new solution, verify it, and request a change for requirements.

\subsubsection{Challenge 9: Difficulty understanding BIM requirements}
Some consultants working on the project are not fluent or familiar with the language in which the requirements documents are written, as reported in Case One by the BIM manager (A3). It is difficult for those consultants to understand the BIM requirements. Although the consultants translate those documents into their language, they are not confident that the translation is accurate.

\textit{Consultants are unfamiliar with the documents' language} was identified as the cause of this issue. ``The language is always a problem when you are an international consultant'' (A3).

\textit{Long time spent at the beginning} is the consequence of this challenge. At the beginning of the project, it took a while for both the client and the contractor to coordinate and make sure that they were on the same page; as mentioned by A3 ``It took some time coordinating with the client to understand the requirements'' (A3).

This challenge is not specified to BIM requirements only. The project has many people (requirements engineers and designers) with different language proficiency, whom may find it difficult to interpret the requirements if the translation is inaccurate.

\subsubsection{Challenge 10: Requirements management tool related}
In Case Two, one of the identified challenges is related to using the requirements management tool DOORS. This challenge was reported by the requirements specialist (C2). People would prefer to use conventional tools like spreadsheets over a specialized new system like DOORS; as mentioned by C2, ``specialists don't really like new systems'' (C2). It is challenging to get people working on the tool. Another part of this challenge is related to reaching models from within DOORS. ``When specialists review stuff, they need to go to the specific document in the specific models to look, and they can't just click on the link, which is annoying'' (C2). The delivered models are verified in a different system, and it could be difficult to find the model and the related requirements during the review process by the client's engineers.

The causes we identified for this challenge are \textit{use of yet another new tool} and  \textit{the limitation of the tool DOORS}. The requirements management tool DOORS is new for many specialists working on the project, and they are not familiar with it; also, some specialists already maintain different systems. ``It is a lot for people that already maintain like fifteen different systems so that was an issue to make them like DOORS'' (C2). Also, DOORS has its limitations, e.g., ``you can't do hyperlinks in the attribute in DOORS'' (C2). Currently, using the tool to link the requirements to a specific attribute or part of the model is not possible.

This challenge leads to \textit{unnecessary work for the requirements specialist} and \textit{extra effort by the people verifying the models}. 
Since DOORS is a new system for people to use, the requirements specialist does spend time preparing views and arranging requirements to make it easy for people involved in the review process. ``I have to be involved quite a bit just for them to know where to look'' (C2). 
Additionally, since the tool has a limitation in linking requirements to the detailed model implementations, model verification requires extra effort. As explained by C2 ``they need to be ready to open several different models which take time to load'' (C2). 

It is important to note here that this challenge does not point out the deficiencies of a specific tool, but rather the lack of integration of the many tools an engineer needs to use to perform their work. 

\subsubsection{Challenge 11: Granularity of traces}
The traces created between the requirements and the models are of high abstraction. This was reported in Case Two by the requirements specialist (C2). There is information used in DOORS to trace requirements to the created models. However, this traceability information is not detailed enough, which makes it difficult to do the verification process. It is challenging to create those kinds of traces as it is seen to be an expensive practice.

\textit{Requirements are not linked to objects} is seen to be as the cause of this challenge, ``even though they give you a specific place to look there will be lots of places to look at'' (C2). The current traces are created between requirements and models rather than requirements and the objects.

One consequence of this challenge is that an \textit{extra effort is required to verify the model} as C2 explained ``it just takes some time and effort'' (C2). Since the requirements are linked to models rather than objects within the model, it takes extra effort and time from the specialist to verify the model. 

\subsubsection{Challenge 12: Lack of Experience Using Tools}

There is a lack of experience in using the modeling tools by the people verifying the design models. This challenge was reported in Case Two by the requirements specialist (C2). The manual model review process, at the client in Case Two, depends mainly on the experience of the specialists. Although, DOORS contains traces between requirements and models, the specialists verifying the model must know where in the model the requirements apply. The specialists lack experience verifying those models, as (C2) put it ``We are not very experienced using the model, so that's a challenge for all specialists'' (C2).

\textit{The use of different models in the project} is the cause of this challenge. There are too many model types used in the project; these models require different tools for viewing. It is difficult for specialists to cope with all these tools. C2 said, ``we can not choose exactly what they are going to use; unfortunately, some consultants use different BIM modeling programs'' (C2). 

\textit{Difficulty navigating the models} is a consequence of lacking the experience in verifying the models. Since the digital assets verification process at the client in Case Two is an experience-based process, it gets difficult to navigate the models to verify them. The specialist needs to figure out where in the model the specific requirements apply.

\subsubsection{Challenge 13: Verifying all requirements}
The client finds it challenging to verify all the requirements in the delivered model. This was reported in Case Two by the requirements specialist (C2). There are many generic and project-specific requirements that apply to the models; it becomes troublesome to verify whether the models conform to all requirements. 

The \textit{absence of risk analysis} for the requirements makes the verification of all requirements difficult. There exists no risk analysis nor classification of requirements based on severity or importance. Therefore, the client has no basis on which to base the prioritization for verification and resource allocation.

As a consequence, there is an \textit{Uncertainty in the verification process} at the client side. Currently, it is not determined if a complete model verification, for all requirements that apply, is necessary or not. ``There is a debate at the client whether or not we are supposed to do a complete verification or just sample verification and see'' (C2).

\section{Discussion}
\label{sec:discussion}

In this section, we discuss the implications of our study results for research and practitioners. We start by discussing the client-supplier communication cycle. Then we look at the challenges and identify potential solution areas from the literature. Finally, we discuss the role of the requirements documents in client-supplier communication.

\subsection{Client-supplier communication}

There are differences between the two cases in the way the requirements are communicated. In Case One, the client documented and released the requirements to the suppliers once, at the beginning of the project (Figure~\ref{fig:rc-case1}), while in Case Two, the requirements specification is a more mature process, as it adopts an iterative approach where the client and the contractor (i.e, supplier) negotiate the requirements before they are stored in a requirements management tool (Figure~\ref{fig:rc-case2}). Early requirements negotiation supports more correct and feasible requirements specification~\cite{Grunbacher2005Negotiation}, and is part of the recipe of the system engineering best practices~\cite{SEBOK2020NeedsAndRequirements}. The requirements process in Case Two follows best practices and is, therefore, superior to the process in Case One.

In Case One, every party involved in the project does their  verification for the delivered digital assets. The technical check, as we see in Figures~\ref{fig:dav-case1} and~\ref{fig:dav-case2}, is a sample check done manually. There are two drawbacks of this check: 1) it is time consuming, and 2) it does not verify all parts of the delivered digital assets. The supplier takes responsibility for the issues that show later in the project due to unverified requirements. A similar observation was made by Makkinga et al.~\cite{makkinga_successful_2018}, where a supplier carried the responsibility of the verification when the contractor did not have enough resources to conduct a complete verification. Consistency checks, which verify whether the design conforms to the general design requirements such as correct use of units of measurement or file formats, are done (semi-)automatically, as shown in Figure~\ref{fig:dav-case1}. We believe that developing a common checklist would have been beneficial in the early identification of potential issues to requirements. Moreover, we speculate that communication between two parties with various levels of domain knowledge could be facilitated by communication brokers~\cite{damian2013role}. 

The life cycle model used in both cases is sequential. A sequential model is defined by INCOSE~\cite{Incose2015Incose} as a systematic approach for the system development process where the system goes through a sequence of steps from goals definition to a complete system. This sequential model is beneficial to use in our cases which are large projects with different parties involved. However, this model also requires verified traceable requirements~\cite{Incose2015Incose}, which is challenging to achieve. As we saw in Section~\ref{sec:results}, the requirements validation may be overlooked (ch3), and the traceability information is not adequate (ch11), which makes checking all the requirements difficult (ch13).

\subsection{Challenges and potential solutions}

\begin{figure*}
    \centering
    \includegraphics[width=1\textwidth]{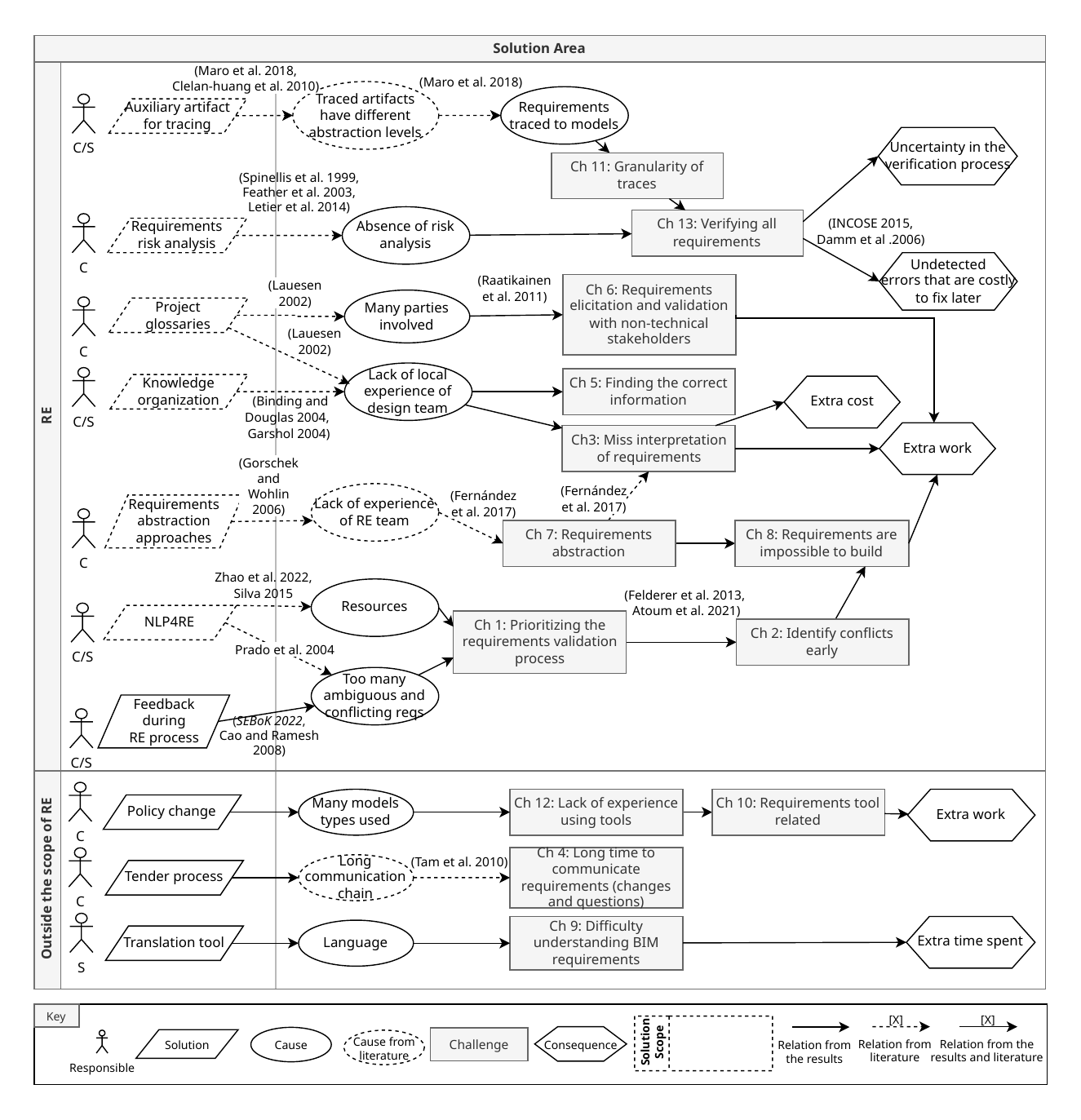}
    \caption{Relationship between challenges}
    \label{fig:challenges}
\end{figure*}

\nocite{damm_faults-slip-throughconcept_2006}
\nocite{raatikainen2011challenges}
\nocite{atoum2021challenges}
\nocite{felderer2013using}
\nocite{tam_impacts_2011}

Figure~\ref{fig:challenges}\footnote{We will make the citations in the figure consistent with the citation in the paper before publishing} depicts the identified challenges, their causes, and their consequences and shows the relations between them based on study results and references from the literature. In addition, we propose solutions areas to tackle these challenges. In the problem analysis, arrows represent a \textit{contributes to} relation between the connected elements. The arrows from the solution area to the problem area indicate whether a solution \textit{potentially addresses} the connected cause and consequently the connected challenge(s). We differentiate between whether the relation originates from:
\begin{itemize}
    \item our results (solid arrow)
    \item an observation made in another study (dashed arrow with reference)
    \item both from our results and from literature (solid arrow with reference) 
\end{itemize}

The solution areas proposed in this section are not exhaustive for all possible solutions for the challenges and their causes. Rather, they are potential solutions where their effectiveness in addressing the challenges has been reported in literature but still needs to be verified in the particular context we have studied. Furthermore, we elaborate on where a solution could be implemented, at the client or the supplier side, by indicating the main \emph{responsible party}. We present the solution areas in more detail next.

\paragraph{Auxiliary artifacts in tracing} When tracing requirements to downstream artifacts in software (e.g., test cases) and system (e.g., design models) projects, an auxiliary artifact could be used to address the different abstraction levels of the traced artifacts. Thus, the auxiliary artifact should be traceable to artifacts with different abstraction levels. It can be an artifact that is produced during the project execution, e.g., a lower level requirement~\cite{maro_software_2018}, or an external artifact that is not part of the software or system, e.g., a domain ontology containing domain concepts and has a hierarchical structure~\cite{cleland-huang_machine_2010}. Using an auxiliary artifact solves the abstraction level mismatch between requirements and the downstream artifacts (e.g., test cases or design models), and consequently could address the granularity of traces challenge (ch11). Fine-grained trace links make it possible to trace requirements to their implementation; thus, verifying all requirements (ch13) in the design model can be more feasible.
This solution should be implemented by the client (for the requirements) and by the supplier (for the delivered artifact).

\paragraph{Requirements risk analysis} A risk analysis could be conducted on all requirements to assess the severity and consequences of failing to fulfill a requirement. A requirements-risk analysis is usually conducted in a similar way as those done for a project plan. In the requirements-risk analysis, three main concepts are identified: problematic requirements, the possible failures, and activities that could prevent or mitigate these risks~\cite{feather_quantitative_2003}. The risk analysis results help engineers in decision-making by reducing uncertainty~\cite{letier_uncertainty_2014}. Such decisions are where resources should be spent~\cite{feather_quantitative_2003} or which requirements should be prioritized for verification~\cite{spinellis_security_1999}.
The client should be responsible for introducing requirements risk analysis as part of the RE process to ensure that the project adheres to the requirements that threaten the project success.

\paragraph{Project glossaries} Project glossaries consist of a set of terms and their descriptions in a specific domain or a project. Project glossaries are used to mitigate misinterpretations of terms (ch3) by project members with different backgrounds~\cite{lauesen2002software}.  Implementing project glossaries can address differences in experience of design teams align terminology between parties with different backgrounds involved during requirements elicitation (ch6).

Project glossaries address these issues by providing an explicit description of terms that can be misunderstood. Delisle and Olson investigate whether project-based terminology and definitions are actually as widely accepted as believed and conclude that more effort should be dedicated towards coordination of glossaries and dissemination of information about project management terms and definitions~\cite{DELISLE2004327}. We speculate that a broader adoption of glossaries can be supported by automated construction of project glossaries~\cite{arora_automated_2017,wang_learning-based_2019}. In our case, the client specifies the project requirements, so it is his responsibility to develop and maintain the project glossary.

\paragraph{Knowledge organization} Knowledge organization is the process of indexing, classifying, and archiving documents and books to ease access to them~\cite{hjorland_what_2008}. McClory et al. stressed the importance of knowledge management and organisational learning and suggested triple-loop learning as an organisational structure~\cite{MCCLORY20171322}.

Infrastructure projects contain many documents (e.g., rules and regulations) that need to be structured and organized to be useful for requirements extraction and analysis. Garshol~\cite{garshol_metadata_2004} identified different types of knowledge organization systems, e.g., controlled vocabularies and taxonomies. The use of an appropriate knowledge organization system to structure domain knowledge~\cite{binding2004kos} could allow engineers to find information about local projects more efficiently and effectively (ch5). This solution could be introduced either by the client or the supplier.  The client owns the rules and regulations, and organizing them through automated classification makes access to information more efficient. The supplier could be obligated to adhere to international standards which apply to multiple projects.

\paragraph{Requirements abstraction methods} First, one of the causes of the major challenges in requirements engineering is the existence of too abstract requirements, which could result in miss interpretation (ch3)~\cite{fernandez_naming_2017}. Thus, a requirement should not be too abstract or unambiguous~\cite{ieee_recommendedpractices_1998}. Second, according to SWEBOK, a requirement is defined as ``a property that must be exhibited by something in order to solve some problem in the real world'' and a good SRS should be an agreement between the client and supplier about ``what the software
product is to do'', not how to do it~\cite{ieee_recommendedpractices_1998}. Thus, a requirement should not restrict possible solutions or be impossible to be realized (ch8). We argue that requirement abstraction methods should be applied to ensure requirements are written on an appropriate abstraction level. Gorschek and Wohlin~\cite{gorschek_requirements_2006} have developed the Requirements Abstraction Model (RAM), a method to systematically specify requirements on multiple levels of abstraction (from product to component level). RAM has been shown to improve the requirements engineering process and the quality of the requirements specified in practice~\cite{gorschek_industry_2007}. Liebel et al. suggested coordinating requirements on various abstraction to avoid communication and coordination breakpoints \cite{liebel2018organisation}. 
Requirements engineers on the client side should adopt one of these solutions as part of the RE process.

\paragraph{NLP4RE} 
Natural language processing (NLP) is the use of (semi-)automated techniques to analyze and model human language~\cite{hirschberg_advances_2015}, mainly through employing machine learning. This is particularly beneficial as the majority of requirements in software and system projects are specified in natural language~\cite{kassab2014state,wagner2019status}. NLP approaches have shown their potential to support the requirements validation process as presented by Zhao et al.~\cite{zhao_natural_2022} in their review of the literature on NLP for requirements engineering (NLP4RE). The use of NLP4RE reduces the number of resources required for the requirements validation~\cite{da_silva_specqua_2015,zhao_natural_2022}, e.g., by performing automated model checking. Moreover, NLP4RE supports requirements engineers in identifying conflicts and ambiguities (ch2), which requires expertise~\cite{prado_leite_perspectives_2004}.
Both the client and supplier could benefit from adopting this solution in the RE process.

\paragraph{Feedback during RE process} One of the good practices in system engineering~\cite{SEBOK2020NeedsAndRequirements} and software engineering~\cite{ramesh_agile_2010} during the requirements engineering process is adopting an iterative approach, where the client and supplier agree on the requirements to be developed incrementally. This approach increases the understanding of requirements and helps with their prioritization~\cite{cao_agile_2008}. In Case Two, an iterative approach for RE is adopted where the client and the contractor (i.e., supplier) negotiate the requirements before they are stored in a requirements management tool (Figure~\ref{fig:rc-case2}). However, in Case One, the requirements were specified and communicated up-front, and changes were difficult to introduce due to contractual obligations. Adopting an iterative approach when specifying the requirements where the feedback of the supplier is considered reduces ambiguity and conflict in the specified requirements.
Both the client and the supplier are responsible for implementing this solution.

\paragraph{Outside the scope of RE} Other challenges (4,9,10,12) and solutions areas are outside the scope of RE but in the area of project management. A change in companies' policy (client), an improved tender process (client), and using better tools are required in order to tackle these challenges.

\paragraph{Observations}
The challenges that we identified in this study and their mapping to potential solutions, as we presented in this section, illustrate the similarity between the system and software engineering domain. Both domains use natural language to specify requirements, and both have digital assets that need to be verified against the specified requirements. In the software domain, engineers produce digital artifacts such as design documents and source code, while in the system domain, engineers produce design models (digital twins of the system) before building the physical build. Therefore, when researchers are looking for a solution to specific challenges in one domain, they should explore existing solutions in the other. Moreover, adopting requirements communication techniques is highly important for the system domain in general and the infrastructure domain in particular, due to the sub-contracted work being the norm, which adds more complexity to the requirements communication process.

\section{Conclusion}
\label{sec:conclusion}

We have conducted two case studies in three companies to explore the requirements validation, requirements communication, digital assets verification processes, and the challenges associated with these processes between the client and the supplier in infrastructure projects. We identified 13 challenges and proposed potential solutions from the literature to address these challenges. Many of the challenges faced, during the forward communication (tender documents) and backward communication (project deliverables), between the client and supplier can be addressed in the area of requirements engineering. Furthermore, the solution for similar challenges in the software domain can potentially address the challenges observed in the system domain. 

The system requirements play a main role in the communication between the client and supplier in infrastructure projects, and their quality likely affects subsequent processes, e.g., verification and acceptance. The verification process is a difficult task in projects on this large scale, with thousands of requirements and models to validate. One particular difficulty is keeping track of the requirements during the verification process.

Future work includes investigating several research areas that we outline below: 

\begin{itemize}
\item Exploring methods to introduce early requirements risk analysis and how to estimate potential risks on early and often incomplete requirements and how to enable risk-based requirements reasoning~\cite{feather_quantitative_2003}.

\item Exploring auxiliary artifacts and their usage scenario in requirements traceability. This involves exploring the relationship between the structural characteristics of auxiliary artifacts (e.g., a taxonomy) and the performance of machine learning models for requirements traceability. 

\item Investigating what types of knowledge organization systems are the most suitable for large-scale infrastructure projects - we believe that project glossaries should be considered as an efficient way of organizing knowledge and building common vocabulary between the parties. We also plan to explore the suitability of topic maps knowledge organizational structure for infrastructure projects~\cite{garshol_metadata_2004}.

\item Investigating the suitability of requirements abstraction models from the software domain (e.g., RAM \cite{gorschek_industry_2007}) for the infrastructure projects. 

\item Exploring the use of NLP4RE techniques to detect conflicts and ambiguity in functional requirements from the infrastructure domain.

\end{itemize}

\section{Acknowledgment}
We write the authors contribution statement following CRediT authorship contribution statement:

\begin{itemize}
    \item Waleed Abdeen: Conceptualization, methodology, formal analysis, investigation, and writing - original draft.
    \item Krzysztof Wnuk: Conceptualization, methodology, investigation, supervision, and writing - review \& editing.
    \item Michael Unterkalmsteiner: Conceptualization, methodology, formal analysis and investigation, supervision, project administration, funding acquisition, and writing - review \& editing
    \item Alexandros Chirtoglou: Conceptualization, formal analysis and writing - review \& editing.
\end{itemize}

The authors declare that they have no known competing financial interests or personal relationships that could have appeared to influence the work reported in this paper. This research was performed within Some Research Project, funded by Some Government Institution, Country.

\bibliography{references}

\end{document}